\documentclass[preprint,aps ,nofootinbib]{revtex4}
\usepackage{}
\usepackage{graphicx}
\usepackage{amsmath}
\usepackage{amsfonts}
\usepackage{amssymb}
\usepackage{color}%
\usepackage{dcolumn}

\providecommand{\U}[1]{\protect\rule{.1in}{.1in}}

\newcommand{\f}{\begin{equation}}
\newcommand{\ff}{\end{equation}}
\newcommand{\fa}{\begin{eqnarray}}
\newcommand{\ffa}{\end{eqnarray}}

\begin{document}

\title{Holographic fermions in charged dilaton black branes}
\author{Wei-Jia Li $^{1}$}
\email{wjli@mail.bnu.edu.cn}
\author{Jian-Pin Wu $^{1}$}
\email{jianpinwu@mail.bnu.edu.cn}
\affiliation{$^1$Department of Physics, Beijing Normal University, Beijing 100875, China}

\begin{abstract}

By imposing the relativistic boundary term and Lorentz violating that
in the dilatonic black brane with a Lifshitz like IR geometry and $AdS_4$ boundary,
we study the properties of the spectral functions of the fermions.
We find that in the two fixed points, there are emergent Fermi-surface structures
and many properties seem to be in agreement with that of Fermi liquid.
Especially, the low energy behavior exhibits a linear dispersion relation.
In addition, we also find that a holographic flat band also emerges
in this background of the dilatonic black brane.

\end{abstract} \maketitle

\section {Introduction}

The physics of many real materials which involve strongly-interacting fermions,
including the cuprate superconductors and other oxides,
seem to lie outside the standard framework of Landau's Fermi-liquid theory.
A general theoretical framework characterizing such new materials remains a suspense.
Recently, many theoretical physicists have resorted to the
AdS/CFT correspondence \cite{Maldacena1997,Gubser1998,Witten1998}
to offer possible insight into such states of matter.
The standard holographic dual for a finite-density, strongly-interacting system is
the Reissner-Nordstr$\ddot{o}$m (RN) black hole in anti-de Sitter (AdS) spacetime.
By using the so-called "bottom-up" approaches, the holographic calculations of
the spectral functions of the fermions in RN black hole
reveal some new classes of non-Fermi liquids \cite{HongLiuNon-Fermi,Sung-Sik Lee,HongLiuAdS2}.
Some extended investigations have also been explored in this background,
including the presence of a magnetic field \cite{Magnetic1,Magnetic2,Magnetic3},
more general fermionic couplings \cite{coupling1,coupling2,coupling3},
the case at finite temperature \cite{FiniteT} etc.
In addition, the extensions to the background of an AdS BTZ black hole, Gauss-Bonnet black hole and the Lifshitz background have also been studied \cite{BTZ,GBJPWu,LifshitzFermions1,LifshitzFermions2}.
On the other hand, some relevant explorations of the spectral function of the holographic fermions have also been performed in a fully consistent "top-down" approach \cite{topdown1,topdown2,topdown3,topdown4}.
For an excellent review on this subjects, we can refer to the lecture \cite{HongLiuLecture}.

However, the background of RNAdS has nonzero ground state entropy density,
which seems to be inconsistent with our intuition that a system of
degenerate fermions has a unique ground state\footnote{Non-degenerate ground states may exist in some topological flat band models, which is not
investigated in this paper\cite{Wang}.}. Therefore, a systematic exploration of the
system that has zero extremal entropy will be important and valuable. Such
models have been proposed in Refs.\cite{ZeroEntropy1,ZeroEntropy2,ZeroEntropy3}.
Especially, in Ref.\cite{FermionsDilatonWu}, the author investigated the low energy behaviors
around the Fermi surface in the background proposed by Gubser and Rocha \cite{ZeroEntropy1}.
They find that the dispersion relation is linear, just like a Fermi liquid.
Furthermore, in Ref.\cite{wywen}, the authors study the dipole coupling effect of holographic fermion in this background. It is found that the linear dispersion relation is robustness for small coupling and a gap emerges the coupling becomes large enough.
To furthermore know the characteristics of the fermionic response
in this kind of background of vanishing ground state entropy density,
in this paper, we will study the low energy behaviors of fermionic response
in another important zero entropy background proposed by Goldstein, etc.\footnote{In the another companion paper\cite{dipoleJPWu}, the dipole coupling effect in this background has been explored.},
which involves a dilaton field coupled to a gauge field.
The near horizon geometry is Lifshitz-like.
It is different from that of the extremal RN black hole,
which is $AdS_{2}\times R^{2}$.

In addition, we also want to study the case of holographic non-relativistic fermionic fixed point
in this background to test the robustness of the emergence of the holographic flat band \cite{HFlatBand}.
By imposing Lorentz violating boundary terms for a spinor field in $AdS_{4}$,
the holographic non-relativistic fermionic fixed points have been implemented in Ref.\cite{HFlatBand},
where a dispersionless flat band emerges. Subsequently, the properties of the spectral function
with bulk dipole coupling have also been studied in Ref.\cite{FlatBandDipole} and in an extremal dilaton black hole geometry in Ref.\cite{FlatBandLMZ}.

The paper is organized as follows. In Section II, we give a brief review
on the extremal charged black brane solutions to Einstein-Maxwell-Dilaton model.
Then, in Section III, we derive the bulk Dirac equations and discuss the holographic calculations of
the retarded Green's functions of those fermionic operators
for relativistic fermionic fixed point and non-relativistic case, respectively.
The numerical results are presented in Section IV.
Conclusions and discussions follow in Section VI.

\section{Extremal charged black brane solutions to Einstein-Maxwell-dilaton model}

\subsection{Einstein-Maxwell-Dilaton model and holographic dictionary}

Following Ref.\cite{ZeroEntropy2}, we can start with the following Einstein-Maxwell-dilaton action\footnote{For the more general holographic models of charged dilatonic black branes, we can also refer to Refs.\cite{dilatonextention1,dilatonextention2,dilatonextention3,dilatonextention4}.}
\begin{equation}
S=\frac{1}{2\kappa^2}\int
d^4x\sqrt{-g}[R-2(\partial\phi)^2-e^{2\alpha\phi}F^2+\frac{6}{L^2}],
\end{equation}
where $R$ is Ricci scalar, $\phi$ is the dilaton field, $F=dA$ is
the field strength, and $L$ denotes the AdS scale. Then the
equations of motion can be obtained by the variation principle as
\begin{equation}
G_{ab}=2\partial_a\phi\partial_b\phi-(\partial\phi)^2g_{ab}+2e^{2\alpha\phi}(F_{ac}{F_b}^c-\frac{1}{4}F^2g_{ab})+\frac{3}{L^2}g_{ab},\label{gr}
\end{equation}
\begin{equation}
\frac{2}{\sqrt{-g}}\partial_a(\sqrt{-g}g^{ab}\partial_b\phi)=\alpha
e^{2\alpha\phi}F^2,\label{kg}
\end{equation}
\begin{equation}
\frac{1}{\sqrt{-g}}\partial_a(\sqrt{-g}e^{2\alpha\phi}F^{ab})=0.\label{em}
\end{equation}
In what follows, we shall focus ourselves onto the electrically
charged black brane solutions, so we take a metric of the form
\begin{equation}
ds^2=-a^2(r)dt^2+\frac{dr^2}{a^2(r)}+b^2(r)[(dx^1)^2+(dx^2)^2],
\end{equation}
and a gauge field of the form
\begin{equation}
e^{2\alpha\phi(r)}F=\frac{Q}{b^2(r)}dt\wedge dr.
\end{equation}
Notice that such a gauge field satisfies Maxwell equation (\ref{em})
automatically. On the other hand, with the above ansatz,
Klein-Gordon equation (\ref{kg}) reads
\begin{equation}
(a^2b^2\phi')'=-\alpha e^{-2\alpha\phi}\frac{Q^2}{b^2},\label{KG}
\end{equation}
where the prime denote the differentiation with respect to the
coordinate $r$. Similarly, Einstein equation reduces to
\begin{equation}
-\frac{a^3}{b^2}[ab'^2+2b(a'b'+ab^{\prime\prime})]=\phi'^2a^4+e^{-2\alpha\phi}\frac{Q^2a^2}{b^4}-\frac{3
a^2}{L^2},\label{tt}
\end{equation}
\begin{equation}
\frac{b'(2ba'+ab')}{ab^2}=\phi'^2-e^{-2\alpha\phi}\frac{Q^2}{b^4a^2}+\frac{3}{a^2L^2},\label{co}
\end{equation}
\begin{equation}
b[b(a'^2+aa^{\prime\prime})+a(2a'b'+ab^{\prime\prime})]=-\phi'^2a^2b^2+e^{-2\alpha\phi}\frac{Q^2}{b^2}+
\frac{3b^2}{L^2}. \label{xx}
\end{equation}
The constraint equation (\ref{co}) can be further simplified as
\begin{equation}
a^2b'^2+\frac{1}{2}(a^2)'(b^2)'=\phi'^2a^2b^2-e^{-2\alpha\phi}\frac{Q^2}{b^2}+\frac{3b^2}{L^2}.\label{CO}
\end{equation}
Combining it with Eq.(\ref{xx}), we obtain
\begin{equation}
(a^2b^2)^{\prime\prime}=\frac{12b^2}{L^2}.\label{cx}
\end{equation}
On the other hand, combining Eq.(\ref{co}) with Eq.(\ref{tt}), we
obtain
\begin{equation}
\frac{b^{\prime\prime}}{b}=-\phi'^2.\label{ct}
\end{equation}
Below we shall employ Eqs.(\ref{KG}), (\ref{CO}), (\ref{cx})
and (\ref{ct}) as our independent equations of motion.

Now assume that the metric solution takes the standard form of
asymptotic AdS, i.e.,
\begin{equation}
a^2=\frac{r^2}{L^2}+\cdot\cdot\cdot,
b^2=\frac{r^2}{L^2}+\cdot\cdot\cdot,
\end{equation}
where the ellipses denote those subdominant terms at large $r$. Then
it follows from equation of motion that the Maxwell and dilaton
fields can be asymptotically expanded as
\begin{equation}
A_t=\mu+\frac{Qe^{-2\alpha\phi_0}}{r}+\cdot\cdot\cdot,\phi=\phi_0-\frac{\alpha
Q\mu}{3r^3}-\frac{\alpha
Q^2e^{-2\alpha\phi_0}}{4r^4}+\cdot\cdot\cdot.
\end{equation}
Next by the holographic dictionary, the bulk gauge field evaluated
at the boundary $\mu$ serves as the source for a conserved charge
$J^t$ associated with a global $U(1)$ symmetry, and the near
boundary data of the dilaton field $\phi_0$ sources a scalar
operator $O$ with the conformal scaling dimension three. To be more
precise, the expectation value of the corresponding boundary quantum
field theory operators $J^t$ and $O$ can be obtained by variations
of the action with respect to the sources, i.e.,
\begin{equation}
\langle J^t\rangle=\frac{\delta S}{\delta
\mu}=-\frac{2Q}{\kappa^2},\langle O\rangle=\frac{\delta
S}{\delta\phi_0}=-\frac{2\alpha Q\mu}{\kappa^2}.
\end{equation}

We would like to conclude this subsection by pointing out that in
the context of AdS/CFT correspondence all other solutions can
actually be obtained by a suitable shift in the dilaton and
rescaling of the coordinates once there is one bulk solution at
hand\footnote{In general relativity, rescaling, as kind of special
diffeomorphism, should not change physics. However, in the context
of AdS/CFT correspondence which we are working on, those solutions
related by rescaling represent distinct physics contents.}. Speaking
specifically, according to the equations of motion, we can firstly
obtain a valid solution with a different $\phi$ by adding a constant
field to the dilaton but with $Qe^{-\alpha\phi}$ fixed. On the other
hand, by rescaling the coordinates as
\begin{equation}
r=\lambda
\tilde{r},t=\frac{\tilde{t}}{\lambda},x^i=\frac{{\tilde{x}}^i}{\bar{\lambda}}\label{rs}
\end{equation}
with $i=1,2$ and $\lambda=\bar{\lambda}$\footnote{Although the
equations of motion allow for the difference between $\lambda$ and
$\bar{\lambda}$, applying AdS/CFT correspondence requires that they
should be equal to each other if the original solution, as inferred
for our purposes, takes the standard asymptotics of AdS.}, the new
solution with $\tilde{Q}=\frac{Q}{\lambda^2}$ is generated as
\begin{equation}
\tilde{a}=\frac{a}{\lambda},\tilde{b}=\frac{b}{\lambda},\tilde{\phi}=\phi.
\end{equation}

In the following, we shall set $L=1$ and $\kappa=1$ for convenience.

\subsection{Analytical scaling solution near the horizon and its numerical AdS completion at infinity}

To find the scaling solution near the horizon $r_h$, let us firstly
introduce the variable $r_{\ast}=r-r_h$ and make the following ansatz for
the behaviors of $a$, $b$, and $\phi$ near the horizon, i.e.,
\begin{equation}
a=C_1r_{\ast}^\gamma,b=C_2r_{\ast}^\beta,\phi=-K\ln r_{\ast}+C_3,
\end{equation}
where $C_1$, $C_2$, $C_3$, $\gamma$, $\beta$,  and $K$ are all
constants\footnote{Without loss of generality, hereafter we shall
require both $C_1$ and $C_2$ to be positive.}. Now plugging such an
ansatz into the equations of motion and manipulating a little
algebra, we have
\begin{equation}
\gamma=1,\beta=\frac{(\frac{\alpha}{2})^2}{1+(\frac{\alpha}{2})^2},K=\frac{\frac{\alpha}{2}}{1+(\frac{\alpha}{2})^2},C_1^2=\frac{6}{(\beta+1)(2\beta+1)},
Q^2e^{-2\alpha C_3}=\frac{(2\beta+1)KC_1^2C_2^4}{\alpha}.
\end{equation}
Whence the metric component $g_{tt}$ has a double zero at the
horizon where $g_{xx}$ also vanishes\footnote{Note that in the case
of $\alpha=0$, such a solution reproduces the near horizon geometry
of the extremal RN black brane. Hereafter we shall assume
$\alpha>0$.}, which means that such a solution corresponds to the
extremal black brane with zero temperature and zero entropy. In
addition, such a solution has a Lifshitz-like symmetry in the
metric, with a dynamical critical exponent $z=\frac{1}{\beta}$,
although such a symmetry is broken by the logarithmic dependence of
the dilaton on $r_{\ast}$.

It is noteworthy that the above solution is an exact solution to the
equations of motion. However, for our purposes it does not have the
expected asymptotic behavior, as we are interested in a solution
which asymptotes to AdS at infinity. As demonstrated below, thanks
to the non-linearity of equations of motion, such a desired solution
can actually be obtained by adding a perturbation to the above
scaling solution near the horizon and evolving it to infinity.
Before proceeding, we would like to rescale the coordinates
appropriately, i.e.,
\begin{equation}
r=\lambda\tilde{r},t=\frac{\tilde{t}}{\lambda},
x^i=\frac{{\tilde{x}}^i}{\lambda'}
\end{equation}
with $\lambda=e^{\frac{C_3}{K}}$ and
$\bar{\lambda}=\sqrt{C_2\lambda^\beta}$, such as to set the constant
$C_2$ to unity and $C_3$ to zero. Whence we have
\begin{equation}
a=C_1r_{\ast},b=r_{\ast}^\beta,\phi=-K\ln r_{\ast},
\end{equation}
and the charge is also fixed in terms of $\alpha$ as
\begin{equation}
Q^2=\frac{6}{\alpha^2+2}.
\end{equation} By
requiring that near the horizon the perturbated functions $a$, $b$,
and $\phi$ solve the equations of motion to leading order, some
tedious but straightforward calculations yield
\begin{equation}
a=C_1 r_{\ast}(1+d_1r_{\ast}^\nu),b=r_{\ast}^\beta(1+d_2r_{\ast}^\nu),\phi=-K\ln r_{\ast}+d_3r_{\ast}^\nu,
\end{equation}
where $C_1$, $\beta$, and $K$ keep unchanged,
$d_3=\frac{2\beta+\nu-1}{2K}d_2$,
$d_1=[\frac{2(1+\beta)(1+2\beta)}{(2\beta+2+\nu)(2\beta+1+\nu)}-1]d_2$,
and $\nu=\frac{1}{2}[-2\beta-1+\sqrt{(2\beta+1)(10\beta+9)}]$ is
positive, implying that the perturbation dies out as the horizon is
approached. Note that such a perturbated solution is characterized
by two parameters, namely, $\alpha$ and $d_1$. For simplicity, we
will focus ourselves onto the case of $\alpha=1$ in the following
discussions.
\begin{figure}
  \includegraphics[]{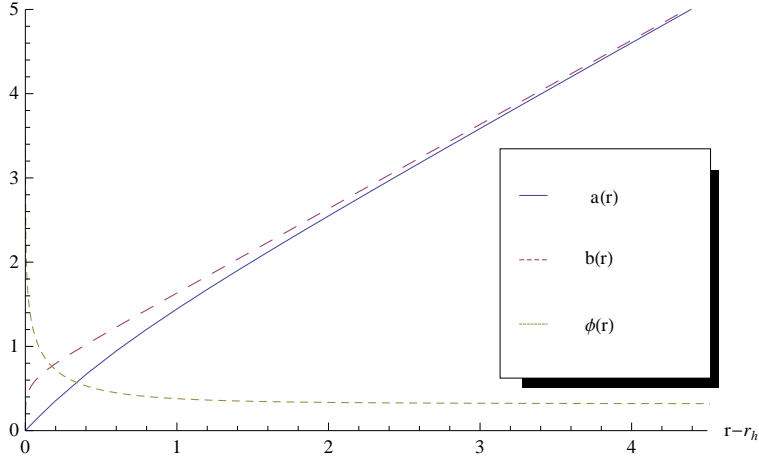}\\
  \caption{Numerical solution interpolating between the near horizon and AdS boundary for $a$, $b$, and $\phi$ in the case of $\alpha=1$ and $d_1=-0.514219$, where the horizon $r_h=0.63539$.}\label{f1}
\end{figure}
\begin{figure}
  \includegraphics[]{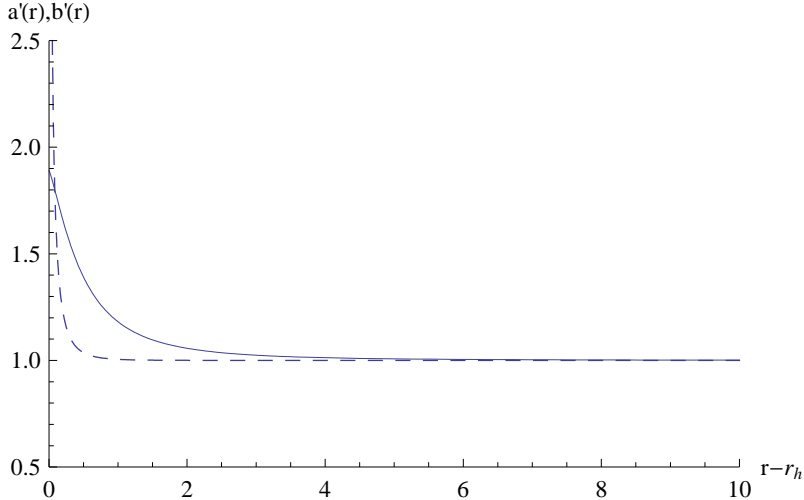}\\
  \caption{Numerical solution interpolating between the near horizon and AdS boundary for $a'$ and $b'$ in the case of $\alpha=1$ and $d_1=-0.514219$, where the solid line denotes
  $a'$ and the dashed line denotes $b'$.}\label{f2}
\end{figure}

Now take the above perturbated solution near the horizon as our
initial data for numerical integration, then we can obtain the
numerical solution to infinity. Generically, adding the perturbation
with $d_1<0$ gives rise to AdS as the conformal boundary while the
numerical solution becomes singular when $d_1>0$. In particular, as
depicted in Figure \ref{f1} and Figure \ref{f2}, for
$d_1=-0.514219$, the dilaton approaches a constant, denoted by
$\phi_0$, and $a=b=r$ for large $r$. So near the boundary the
solution takes the standard asymptotics of AdS\footnote{As pointed
out before, with such a solution at hand, we can actually generate
any other solution in the context of AdS/CFT correspondence. In
particular, the solution for any other negative $d_1$ can be
obtained by rescaling as Eq.(\ref{rs}) with
$\lambda=(-\frac{d_1}{0.514219})^{\frac{1}{\nu}}$ and adding
$K\ln\lambda$ to the dilaton with an additional rescaling as
$x^i=\sqrt{\lambda^{1-\beta}}{x'}^i$. This additional rescaling
implies that $b=\lambda^{1-\beta}r$ for large $r$. In other words,
for any other negative $d_1$, $b$ does not take the standard form of
asymptotic AdS.}.

In the next section, we shall investigate the holographic fermions
in such a background, where without loss of generality, we shall
take $Q=\sqrt{2}$, from which we can numerically obtain a negative
chemical potential $\mu\simeq-0.843$ for the boundary theory.

\section{Holographic fermions in extremal
charged dilaton black branes\label{mr}}

\subsection{Bulk equation of motion}

Considering the following bulk fermion action
\begin{eqnarray}
\label{actionspinor}
S=i\int d^{d+1}x \sqrt{-g}\overline{\zeta}\left[\frac{1}{2}(\overrightarrow{\mathcal {D}_a}-\overleftarrow{\mathcal {D}_a})-m\right]\zeta+S_{bd},
\end{eqnarray}
where $\Gamma^{a}$ is related to the usual
flat space gamma matrix by a factor of the vielbein,
$\Gamma^{a}=(e_{\mu})^{a}\Gamma^{\mu}$, $\overrightarrow{\mathcal{D}_{a}}=\Gamma^{a}[\partial_{a}+\frac{1}{4}(\omega_{\mu\nu})_{a}\Gamma^{\mu\nu}-iqA_{a}]$
is the covariant derivative with $(\omega_{\mu\nu})_{a}$ the spin connection 1-forms, and the boundary term $S_{bd}$ should be added to have a well-defined variational principle for the action, and we will give a detailed discussion in the next subsection.
The Dirac equation derived from the action $S$ is expressed as
\begin{eqnarray}
\label{DiracEquation1}
(\overrightarrow{\mathcal{D}_{a}}-m)\zeta=0.
\end{eqnarray}

Making a transformation
$\zeta=(-g g^{rr})^{-\frac{1}{4}}\mathcal{F}$ to remove the spin
connection and expanding $\mathcal{F}$ as $\mathcal{F}=\tilde{F} e^{-i\omega t +ik_{i}x^{i}}$
in Fourier space, the Dirac equation (\ref{DiracEquation1}) turns out to be
\begin{eqnarray}
\label{DiracEinFourier}
\sqrt{g^{rr}}\Gamma^{r}\partial_{r}\tilde{F}
-i(\omega+q A_{t})\sqrt{g^{tt}}\Gamma^{t}\tilde{F}
+i k \sqrt{g^{xx}}\Gamma^{x}\tilde{F}
-m \tilde{F}=0.
\end{eqnarray}
where due to rotational symmetry in the spatial directions,
we set $k_{x}=k$ and $k_{x}\neq 0,~i\neq x$ without losing generality.
Notice that Eq. (\ref{DiracEinFourier}) only depends on three Gamma matrices $\Gamma^{r},\Gamma^{t},\Gamma^{x}$.
So it is convenient to split the spinors $\tilde{F}$ into $\tilde{F}=(F_{1},F_{2})^{T}$ and
choose the following basis for our gamma matrices as in \cite{HongLiuAdS2}:
\begin{eqnarray}
\label{GammaMatrices}
 && \Gamma^{r} = \left( \begin{array}{cc}
-\sigma^3   & 0  \\
0 & -\sigma^3
\end{array} \right), \;\;
 \Gamma^{t} = \left( \begin{array}{cc}
 i \sigma^1    & 0  \\
0 & i \sigma^1
\end{array} \right),  \;\;
\Gamma^{x} = \left( \begin{array}{cc}
-\sigma^2    & 0  \\
0 & \sigma^2
\end{array} \right),
\qquad \ldots
\end{eqnarray}

Then, we have a new version of the Dirac equation as
\begin{eqnarray} \label{DiracEF}
\sqrt{g^{rr}}\partial_{r}\left( \begin{matrix} F_{1} \cr  F_{2} \end{matrix}\right)
+m\sigma^3\otimes\left( \begin{matrix} F_{1} \cr  F_{2} \end{matrix}\right)
=\sqrt{g^{tt}}(\omega+qA_{t})i\sigma^2\otimes\left( \begin{matrix} F_{1} \cr  F_{2} \end{matrix}\right)
\mp  k \sqrt{g^{xx}}\sigma^1 \otimes \left( \begin{matrix} F_{1} \cr  F_{2} \end{matrix}\right)
~.
\end{eqnarray}

Furthermore, according to eigenvalues of $\Gamma^{r}$,
we make such a decomposition $F_{\pm}=\frac{1}{2}(1\pm \Gamma^{r})\tilde{F}$. Then
\begin{eqnarray} \label{gammarDecompose}
F_{+}=\left( \begin{matrix} 0\cr \mathcal{B}_{1} \cr  0\cr \mathcal{B}_{2} \end{matrix}\right),~~~~
F_{-}=\left( \begin{matrix} \mathcal{A}_{1} \cr  0\cr \mathcal{A}_{2}\cr 0 \end{matrix}\right),~~~~
with~~~~F_{\alpha} \equiv \left( \begin{matrix} \mathcal{A}_{\alpha} \cr  \mathcal{B}_{\alpha} \end{matrix}\right),\alpha=1,2.
\end{eqnarray}

Under such decomposition, the Dirac equation (\ref{DiracEF}) can be rewritten as
\begin{eqnarray} \label{DiracEAB1}
(\sqrt{g^{rr}}\partial_{r}\pm m)\left( \begin{matrix} \mathcal{A}_{1} \cr  \mathcal{B}_{1} \end{matrix}\right)
=\pm(\omega+qA_{t})\sqrt{g^{tt}}\left( \begin{matrix} \mathcal{B}_{1} \cr  \mathcal{A}_{1} \end{matrix}\right)
-k \sqrt{g^{xx}} \left( \begin{matrix} \mathcal{B}_{1} \cr  \mathcal{A}_{1} \end{matrix}\right)
~,
\end{eqnarray}
\begin{eqnarray} \label{DiracEAB2}
(\sqrt{g^{rr}}\partial_{r}\pm m)\left( \begin{matrix} \mathcal{A}_{2} \cr  \mathcal{B}_{2} \end{matrix}\right)
=\pm(\omega+qA_{t})\sqrt{g^{tt}}\left( \begin{matrix} \mathcal{B}_{2} \cr  \mathcal{A}_{2} \end{matrix}\right)
+k \sqrt{g^{xx}} \left( \begin{matrix} \mathcal{B}_{2} \cr  \mathcal{A}_{2} \end{matrix}\right)
~.
\end{eqnarray}

Introducing the ratio $\xi_{\alpha}\equiv \frac{\mathcal{A}_{\alpha}}{\mathcal{B}_{\alpha}}$,
one can package the Dirac equations (\ref{DiracEAB1}) and (\ref{DiracEAB2})
into the evolution equation of $\xi_{\alpha}$,
\begin{eqnarray} \label{DiracEF1}
(\sqrt{g^{rr}}\partial_{r}
+2m)\xi_{\alpha}
=\left[ \sqrt{g^{tt}}(\omega+q A_{t})+ (-1)^{\alpha} k \sqrt{g^{xx}}  \right]
+ \left[ \sqrt{g^{tt}}(\omega+q A_{t})- (-1)^{\alpha} k \sqrt{g^{xx}}  \right]\xi_{\alpha}^{2}
~,
\end{eqnarray}
which will be more convenient to impose the boundary conditions at the horizon
and read off the boundary Green functions.

Now substitute the near horizon metric
into Eq. (\ref{DiracEF}), we wind up with
\begin{eqnarray} \label{GatTip1}
\partial_{r_*}\left( \begin{matrix} F_{1} \cr  F_{2} \end{matrix}\right)
+\frac{\sqrt{7}m}{5r_*}\sigma^3\otimes\left( \begin{matrix} F_{1} \cr  F_{2} \end{matrix}\right)=\frac{7}{25r_*^2}(\omega-\frac{5\sqrt{2}}{7}qr_*^{\frac{7}{5}})i\sigma^2\otimes\left( \begin{matrix} F_{1} \cr  F_{2} \end{matrix}\right)\mp  \frac{\sqrt{7}}{5r_*^{\frac{6}{5}}}k\sigma^1 \otimes \left( \begin{matrix} F_{1} \cr  F_{2} \end{matrix}\right)
\end{eqnarray}
for the equation of motion and
\begin{eqnarray} \label{GatTip2}
\partial_{r_*}\xi_\alpha
+\frac{2\sqrt{7}m}{5r_*}\xi_\alpha&=&\Big[\frac{7}{25r_*^2}(\omega-\frac{5\sqrt{2}}{7}qr_*^{\frac{7}{5}})+(-1)^\alpha  \frac{\sqrt{7}}{5r_*^{\frac{6}{5}}}k\sigma^1\Big]\nonumber\\&+&\Big[\frac{7}{25r_*^2}(\omega-\frac{5\sqrt{2}}{7}qr_*^{\frac{7}{5}})-(-1)^\alpha  \frac{\sqrt{7}}{5r_*^{\frac{6}{5}}}k\sigma^1\Big]\xi_\alpha^2.
\end{eqnarray}
Whence the in-falling boundary condition near the horizon is
\begin{eqnarray} \label{GatTip3}
F_\alpha \propto \left(
                   \begin{array}{c}
                     i \\
                     1 \\
                   \end{array}
                 \right)e^{-i\omega \mathcal {R}}
,~~~~for~~\omega\neq 0.
\end{eqnarray}
and \begin{eqnarray} \label{GatTip4}
F_\alpha \propto \left(
                   \begin{array}{c}
                     |k| \\
                    (-1)^\alpha k \\
                   \end{array}
                 \right)e^{|k| \mathcal {\bar{R}}}
,~~~~for~~\omega= 0,
\end{eqnarray}
where $\mathcal {R}=\frac{7}{25}\int\frac{dr_*}{r_*^2}$ and $\mathcal {\bar{R}}=\frac{\sqrt{7}}{5}\int\frac{dr_*}{r_*^{6/5}}$.
Thereby, the boundary conditions for $\
\xi_\alpha$ are
\begin{eqnarray} \label{GatTip}
\xi_{\alpha}\buildrel{r \to r_{h}}\over =i,~~~~for~~\omega\neq 0.
\end{eqnarray}
and \begin{eqnarray} \label{GatTip3}
\xi_{\alpha}\buildrel{r \to r_{h}}\over =(-1)^\alpha \textbf{sign}(k),~~~~for~~\omega=0.
\end{eqnarray}
In the subsequent subsection, we will discuss how to read off the boundary Green functions
for two different fermionic fixed points, respectively.

\subsection{Boundary terms and fermionic fixed points}

\subsubsection{Relativistic fermionic fixed point}

So far, most of work on the holographic fermionic systems focuses on
the perturbations on the relativistic fixed point.
Taking the following boundary term can keep Lorentz invariance for the boundary theory
\begin{eqnarray} \label{Boundaryterm1}
S_{bd}=\pm\frac{i}{2}\int\sqrt{-gg^{rr}}\bar{\zeta}\zeta
\end{eqnarray}
For $+$ sign, varying the on-shell action, we obtain:
\begin{eqnarray} \label{variation}
\delta S&=&i\int d^3x (\delta \bar{F_+}F_- + \bar{F_-}\delta F_+)\nonumber\\
&=&-\int d^3x (\delta\mathcal{B}_{1}^\dag\mathcal{A}_{1}+\delta\mathcal{B}_{2}^\dag
\mathcal{A}_{2}+\mathcal{A}_{1}^\dag\delta\mathcal{B}_{1}+\mathcal{A}_{2}^\dag\delta\mathcal{B}_{2})
\end{eqnarray}
If we fix $\mathcal{B}_{1}$ and $\mathcal{B}_{2}$, i.e. impose Dirichlet boundary conditions for $F_+$, this choice of
boundary condition is usually referred to as the standard quantization for fermions. The dimension of the
boundary fermionic operator is $\Delta_+=\frac{3}{2}+m$. Inversely, for $-$ sign, if we impose Dirichlet boundary conditions for $F_-$, this choice of
boundary condition is usually referred to as the alternative quantization. The dimension of the
boundary fermionic operator is $\Delta_-=\frac{3}{2}-m$. For the special case $m=0$, these two CFTs are equivalent.

Near the boundary, a solution of the Dirac equation (\ref{DiracEF}) can be expressed as
\begin{eqnarray} \label{BoundaryBehaviour}
F_{\alpha} \buildrel{r \to \infty}\over {\approx} a_{\alpha}r^{m}\left( \begin{matrix} 0 \cr  1 \end{matrix}\right)
+b_{\alpha}r^{-m}\left( \begin{matrix} 1 \cr  0 \end{matrix}\right),
\qquad
\alpha = 1,2~.
\end{eqnarray}

If $b_{\alpha}\left( \begin{matrix} 1 \cr  0 \end{matrix}\right)$
and $a_{\alpha}\left( \begin{matrix} 0 \cr  1 \end{matrix}\right)$ are related by
$b_{\alpha}\left( \begin{matrix} 1 \cr  0 \end{matrix}\right)
=\mathcal{S}a_{\alpha}\left( \begin{matrix} 0 \cr  1 \end{matrix}\right)$,
then the boundary Green's functions $G$ is given by $G=-i \mathcal{S}\gamma^{0}$ \cite{HongLiuSpinor}.
Therefore
\begin{eqnarray} \label{GreenFBoundary}
G (\omega,k)= \lim_{r\rightarrow \infty} r^{2m}
\left( \begin{array}{cc}
\xi_{1}   & 0  \\
0  & \xi_{2} \end{array} \right)  \ ,
\end{eqnarray}

\subsubsection{Non-relativistic fermionic fixed point}

As is pointed in \cite{HFlatBand}, if we take the following boundary term, non-relativistic fixed points can emerge
\begin{eqnarray} \label{boundary2}
S_{bd}&=&\frac{1}{2}\int_{\partial\mathcal{M}}d^3x\sqrt{-gg^{rr}}\bar{\zeta}\Gamma^1\Gamma^2\zeta
\end{eqnarray}
Such a boundary term corresponds to a complex double trace operator with dimension $\Delta=3-m$ on the boundary.
For the case of $m=0$, the operator is marginal and sweeps out a
manifold of fixed points on which there is no Lorentz invariance.
The variation of the on-shell action can been written as
\begin{eqnarray} \label{variation2}
\delta S=-\int d^3x(\delta B_1^\dag A_1+ B_2^\dag \delta A_2+A_1^\dag \delta B_1+ \delta A_2^\dag  B_2)
\end{eqnarray}
where we have defined $(A_1,A_2)=\frac{1}{\sqrt{2}}(\mathcal {A}_1+\mathcal {A}_2,\mathcal {A}_1-\mathcal {A}_2)$ and
$(B_1,B_2)=\frac{1}{\sqrt{2}}(\mathcal {B}_1+\mathcal {B}_2,\mathcal {B}_2-\mathcal {B}_1)$. Thereby, if we impose
Dirichlet boundary conditions for $\left(
                                     \begin{array}{cc}
                                       B_1 & A_2 \\
                                     \end{array}
                                   \right)^T
$ we can derive
\begin{eqnarray} \label{smatrix2}
\left(
  \begin{array}{c}
    A_1 \\
    B_2 \\
  \end{array}
\right)=\mathcal {S}\left(
                      \begin{array}{c}
                        B_1 \\
                        A_2 \\
                      \end{array}
                    \right)
\end{eqnarray}
and the retarded Green function is $G_R=-\mathcal {S}$. According to \cite{HFlatBand,FlatBandDipole},
the retarded function on this fixed point has
a relation between the one on the relativistic fixed point as follows
\begin{eqnarray} \label{green2}
G_{\pm}=\frac{G_{11}G_{22}\pm\sqrt{1+G_{11}^2+G_{22}^2+G_{11}^2G_{22}^2}}{G_{11}+G_{22}}
\end{eqnarray}
For the special case $m=0$, we have $\xi_1=G_{11}=-\frac{1}{G_{22}}$.
Hence,
\begin{eqnarray} \label{smatrix2}
G_{\pm}=\frac{\xi_1\mp1}{1\pm\xi_1}.
\end{eqnarray}
Thereby, we can numerically resolve the equation of $\xi_1$ and
read the response through the above equation.

\section{Numerical results}

\subsection{Relativistic fermionic fixed point}

\subsubsection{General behavior of spectral functions}

\begin{figure}
\center{
\includegraphics[scale=0.9]{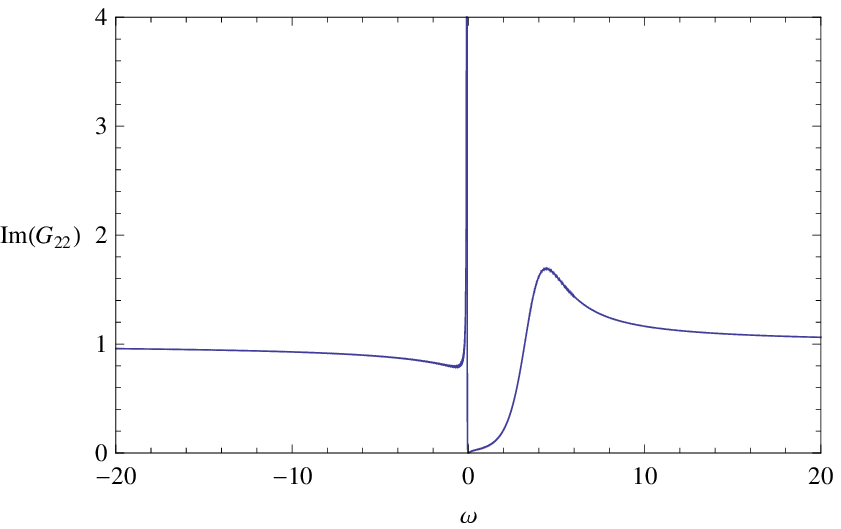}\hspace{1cm}
\includegraphics[scale=0.9]{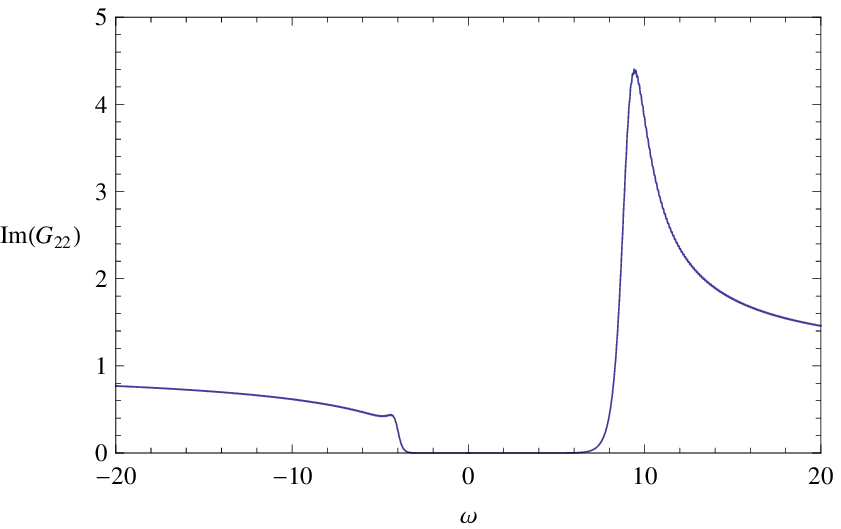}\\ \hspace{1cm}
\caption{\label{GeneralB}Spectral function $ImG_{22}(\omega)$ at $k=1.0< |\mu_{q}|$ (left plot)
and $k=6.0 > |\mu_{q}|$ (right plot) for $m=0$ and $q=4$ ($\mu_{q}\approx -3.37$).
Right plot: $Im G_{22}$ is roughly zero in the rage $\omega\in (-k-\mu_{q},k-\mu_{q})$.}}
\end{figure}

From the evolution equation of $\xi_{\alpha}$ (Eq.(\ref{DiracEF1})),
one can see that the Green function possesses the following symmetry properties:

(1) $G_{22}(\omega,k)=G_{11}(\omega,-k)$;~~~~(2) $G_{22}(\omega,k;-q)=G_{11}^{\star}(-\omega,k;q)$;~~~~

For the case $m=0$,

(3) $G_{22}(\omega,k)=-\frac{1}{G_{11}(\omega,k)}$;~~~~(4) $G_{22}(\omega,k=0)=G_{11}(\omega,k=0)=i$.

Thank to the above symmetry properties, we will focus mainly on $G_{22}$
and restrict ourselves to positive $k$ and $q$ below.
When the background geometry is pure $AdS_{4}$, the Green function (massless bulk fermion)
can be easily obtained as \cite{GreenFpureAdS1,GreenFpureAdS2}
\begin{eqnarray} \label{GreenFpureAdS}
G_{11}=\sqrt{\frac{k+(\omega+i\epsilon)}{k-(\omega+i\epsilon)}},~~~G_{22}=-\sqrt{\frac{k-(\omega+i\epsilon)}{k+(\omega+i\epsilon)}}.
\end{eqnarray}
where $\epsilon\rightarrow 0$. It is clear that the spectral function has a particle-hole symmetry (symmetry under $(\omega,k)\rightarrow (-\omega,-k)$)
and has an edge-singularity along $\omega=\pm k$. In addition, both components of $Im G$ are zero in the region $\omega\in (-k,k)$.

Here, for the charged dilaton black branes background,
we will also do several consistency checks on our numerics as Ref.\cite{HongLiuNon-Fermi,GBJPWu}.
For concreteness, in this paper, we will mainly consider the specific example: $q=4$ and $m=0$.
The dependence of the parameter $q$ and $m$ will be discussed elsewhere.
In FIG.\ref{GeneralB}, we show the spectral function $Im G_{22}$ at $k=1.0<|\mu_{q}|$ (left plot)
and $k=6.0>|\mu_{q}|$ (right plot) for $m=0$ and $q=4$ ($\mu_{q}\approx -3.37$).
Firstly, both of them are asymptote to $1$ as $|\omega|\rightarrow \infty$,
which recovers the behavior in the vacuum.
In addition, for a fixed large $k\gg |\mu_{q}|$,
$Im G_{22}$ is roughly zero in the rage $\omega\in (-k-\mu_{q},k-\mu_{q})$
(right plot in Fig.\ref{GeneralB}). It is one of the features of the vacuum behavior.
However, for $k=1.0<|\mu_{q}|$ (left plot in Fig.\ref{GeneralB}),
the deviation from the vacuum behavior becomes significant.
After having done consistency checks, we will turn to the exploration on some
specific properties of the spectral function in the the charged dilaton black branes background.

\subsubsection{Some specific properties of the spectral functions}\label{SubS1}

\begin{figure}
\center{
\includegraphics[scale=0.6]{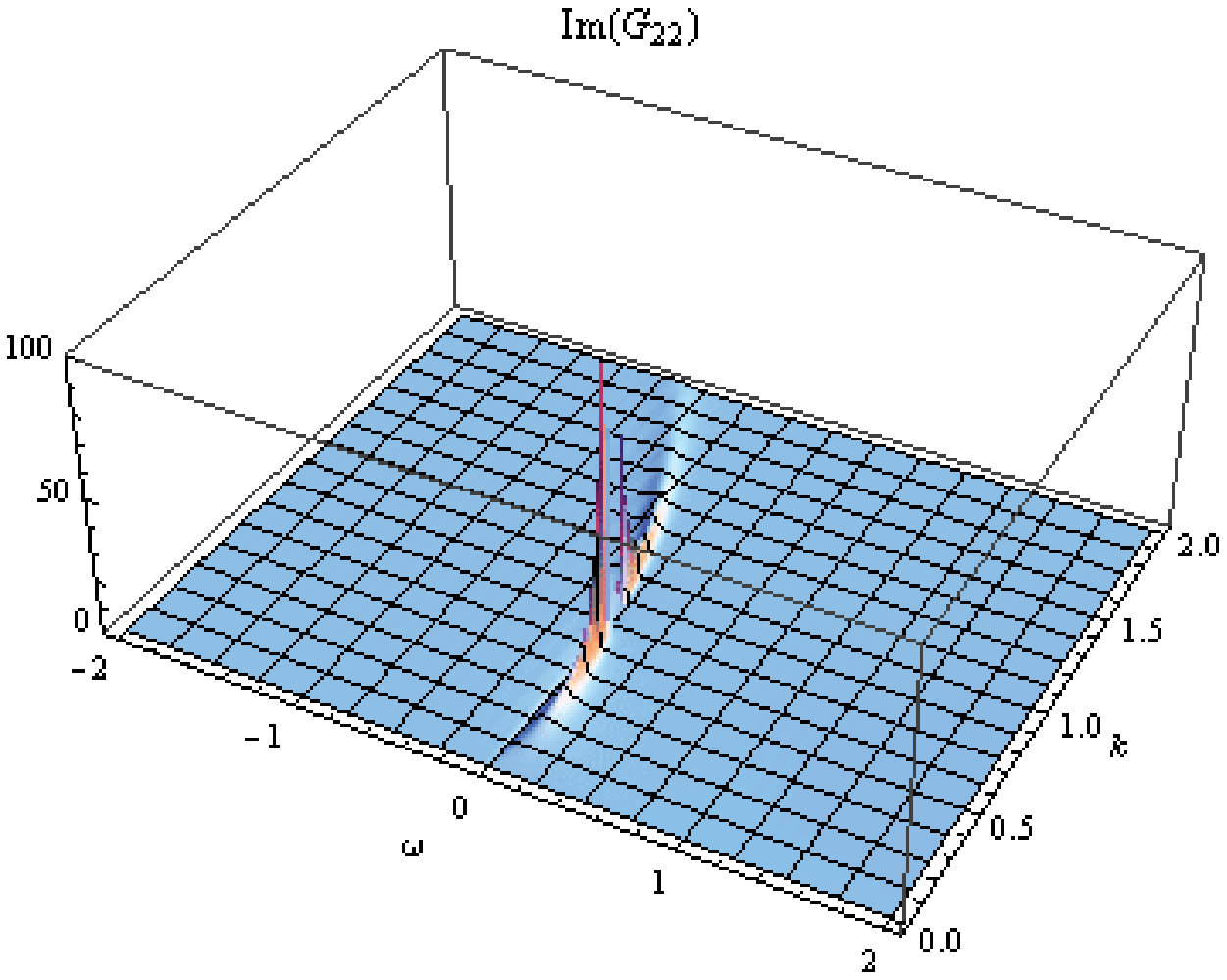}\hspace{1cm}
\includegraphics[scale=0.6]{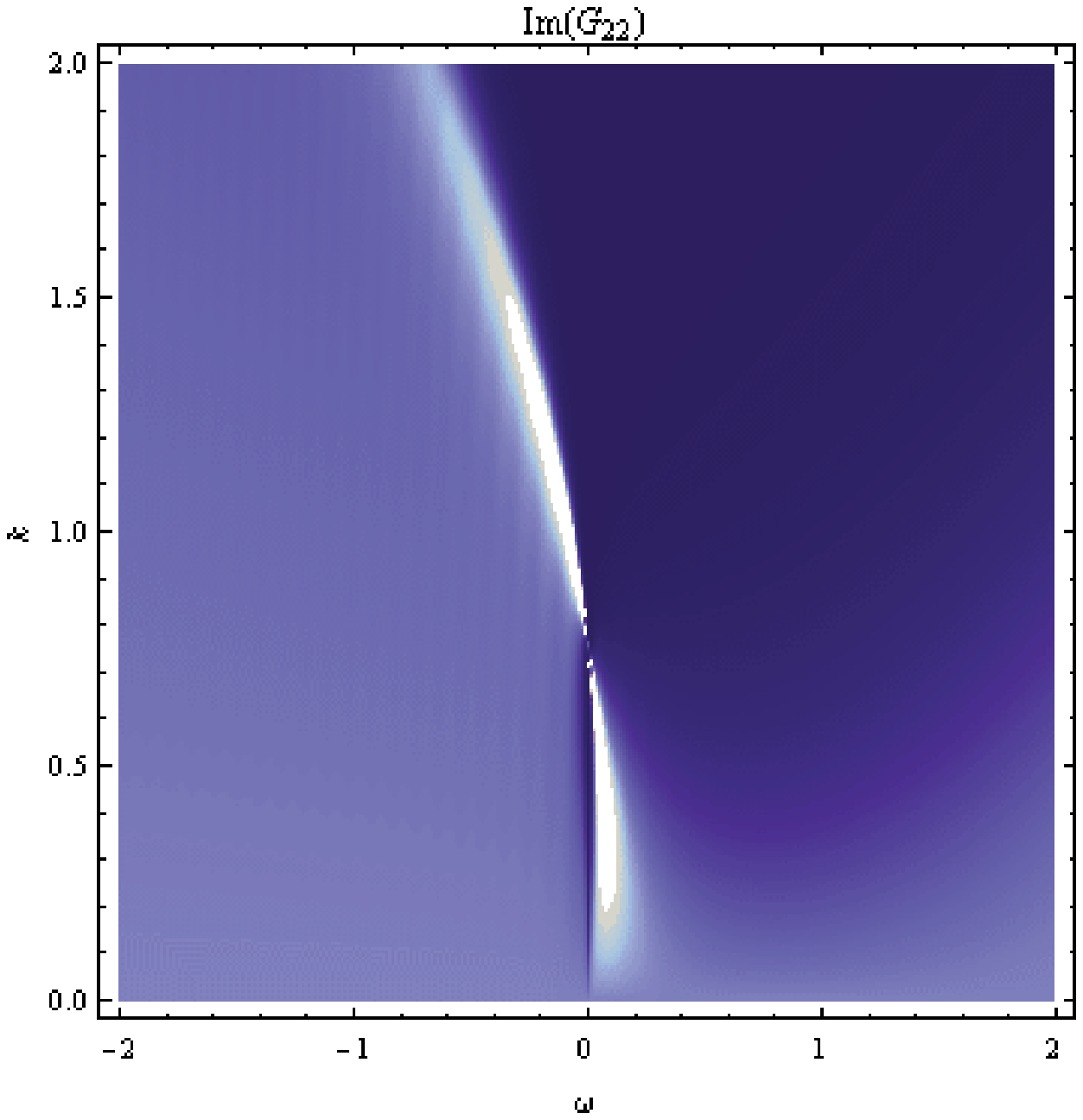}\\ \hspace{1cm}
\caption{\label{FermikF3DDq4}The 3d and density plots of $ImG_{22}(\omega,k)$
for $m=0$ and $q=4$. A sharp quasiparticle-like
peak occurs near $k_{F}\approx 0.75$, indicating a Fermi surface.
Improving the accuracy, the Fermi momentum can be furthermore determined as
$k_{F}\approx 0.756545$.}}
\end{figure}

As shown in the above subsection, the deviation from the vacuum behavior becomes more and more significant
as the decrease of $k$ and near $k_{F}=0.756545$, a sharp quasiparticlelike peak occurs
instead of the finite peak of $ImG_{22}$ in the large $k$ region (Fig.\ref{FermikF3DDq4})\footnote{Fig.\ref{FermikF3DDq4} is took from the plots above in Fig.2 in Ref.\cite{dipoleJPWu}.}.

\begin{figure}
\center{
\includegraphics[scale=1]{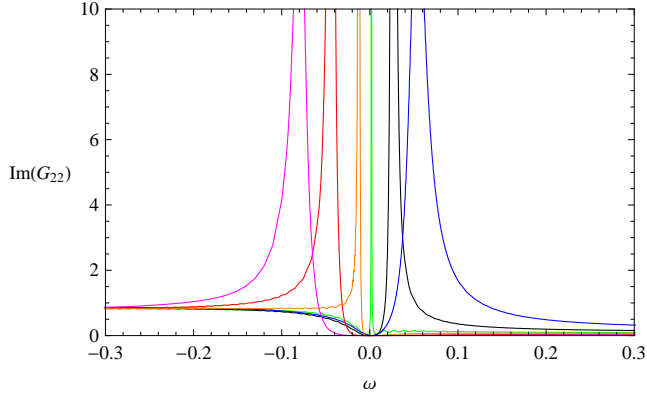}\\ \hspace{1cm}
\caption{\label{detaq4}The plot of $Im G_{22}(k)$ for different $k$.
They show that the quasiparticle peak approach a delta function at the Fermi momentum $k=k_{F}$.
Magenta for $k=1$, red for $k=0.9$, orange for $k=0.8$,
green for $k=0.75$, black for $k=0.65$ and blue for $k=0.5$.}}
\end{figure}
\begin{figure}
\center{
\includegraphics[scale=0.6]{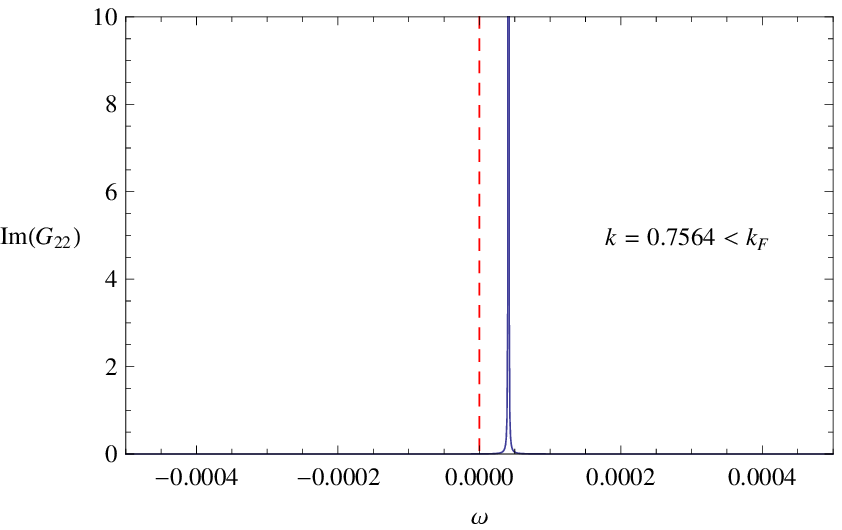}\hspace{0.1cm}
\includegraphics[scale=0.6]{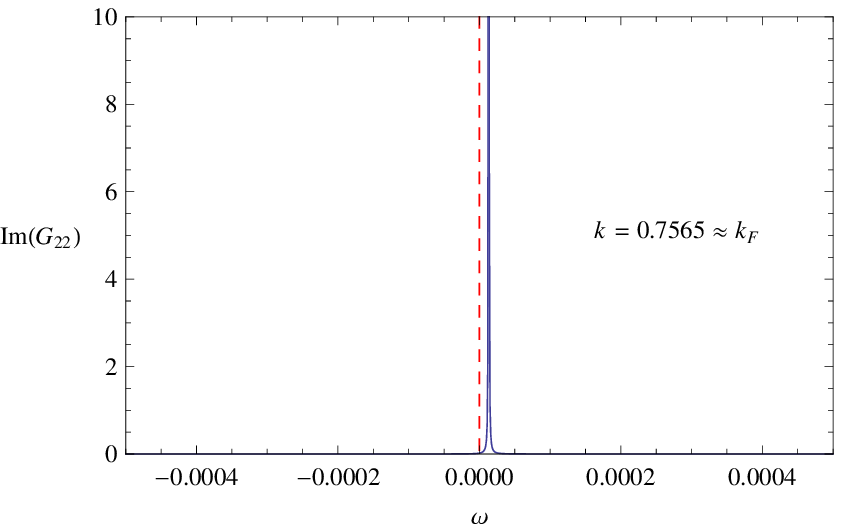}\hspace{0.1cm}
\includegraphics[scale=0.6]{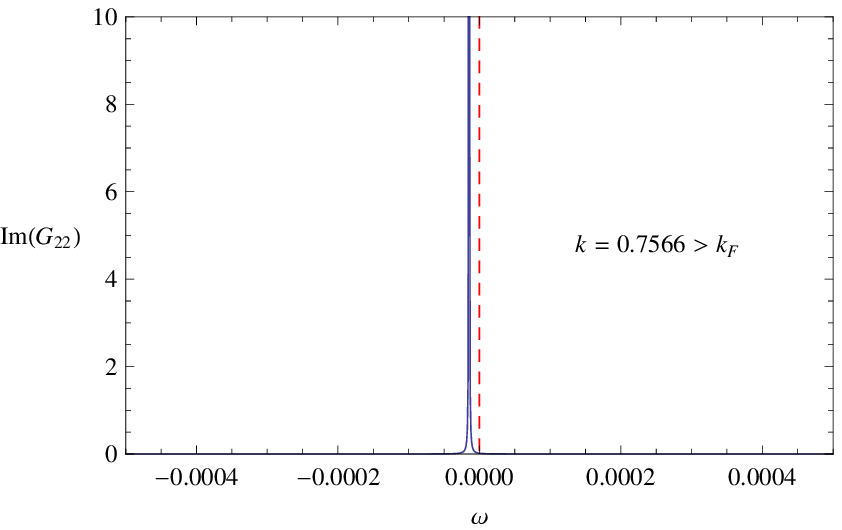}\\ \hspace{0.1cm}
\caption{\label{vanishingG}The plot of $Im G_{22}(k)$ for different $k$.
They show that independent of $k$, the Green's function vanishes at the Fermi energy.}}
\end{figure}

Now, we can move on to explore some specific properties of the spectral functions.
Firstly, from Fig.\ref{detaq4}, one can see that the peak becomes narrower and narrower
when we dial $k$ ($k<k_{F}$) from small to large. When $k$ approaches $k_{F}$,
their heights approach infinity and their widths approach zero, which is almost a delta function.
However, when the Fermi surface is crossed, the peak become wide again.
In addition, independent of $k$, the Green's function vanishes at the Fermi energy
($ImG_{22}(\omega=0,k)=0$). It is shown in Fig.\ref{vanishingG}.

\begin{figure}
\center{
\includegraphics[scale=0.9]{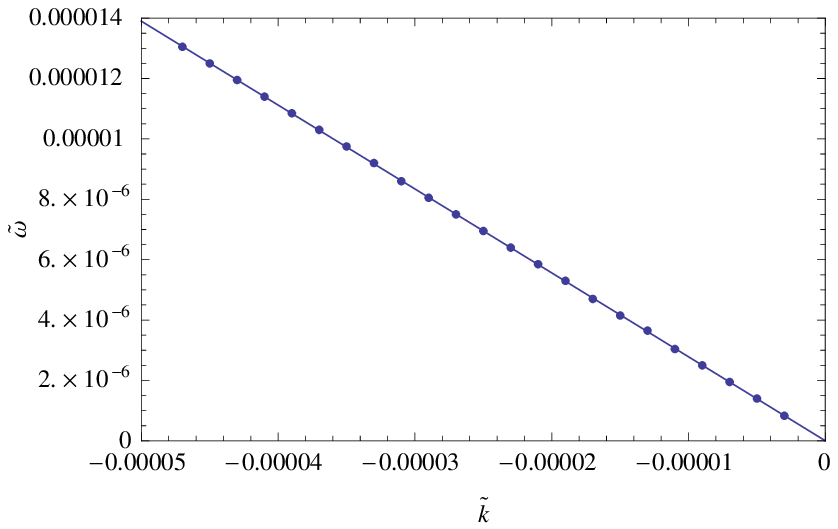}\\ \hspace{1cm}
\caption{\label{DRHSq4}The dispersion relation between $\tilde{k}$ and $\tilde{\omega}$ for the case of relativistic fermionic fixed point.}}
\end{figure}

Especially, the most important thing is the behavior of the spectral function
in the region of small $\tilde{k}=k-k_{F}$ and $\omega$.
By fitting the data, we find that there exists a linear dispersion relation between small $\tilde{k}$ and $\tilde{\omega}(\tilde{k})$ (Fig.\ref{DRHSq4}): $\tilde{\omega}(\tilde{k})\sim \tilde{k}$,
where $\tilde{\omega}$ is the location of the maximum of the peak.
Here we would like to stop and give some comments.
The three characteristics above, especially the linear dispersion relation,
seem to indicate that this fermionic system in extremal charged dilaton black branes is a Fermi liquid
\footnote{The another important scaling behavior of the height of $ImG_{22}$ at the maximum is hard
to fit well numerically, we will expect to explore it by performing the analytical approximation method.}.

Finally, we would like to point out that by using analytic method, the structure of the fermion response in the Einstein-Maxwell-dilaton black holes \cite{ZeroEntropy2} was worked out already in Ref.\cite{dilatonextention4}.
They find that for $\beta+\gamma> 1$, which includes the case we investigate in this paper
\footnote{In this paper, we set $\gamma=\alpha=1$, so that $\beta=1/5$,
which obviously belongs to $\beta+\gamma> 1$.},
the Green's function is
\begin{eqnarray} \label{GR22A}
G_{22}=\frac{c_{3}}{\omega-v_{F}(|k|-k_{F})+ic_{2}e^{-2I}},
\end{eqnarray}
with
\begin{eqnarray} \label{I}
I=c_{1}\left(\frac{k^{2\gamma-1}}{\omega^{\beta+\gamma-1}}\right)^{\frac{1}{\gamma-\beta}},
\end{eqnarray}
where $c_{2}$ and $c_{3}$ are real constant and $c_{1}$ is positive constant. When $\omega\rightarrow 0$,
$e^{-2I}$ is exponentially suppressed. Therefore, from the analytic expression of the Green's function (\ref{GR22A}),
we find that the dispersion relation is linear, which is consistent with our result.

\subsection{Non-relativistic fermionic fixed point}

Now, we will turn to explore the non-relativistic fermionic fixed point case.
As revealed in Ref.\cite{HFlatBand}, a holographic flat band can emerge
in the case of non-relativistic fermionic fixed point.
Here we want to know whether the emergence of the flat band is robust in the dilatonic background.
As the case of relativistic fermionic fixed point, we will also mainly focus on the case of $q=4$.
From Fig.\ref{FlatBand3DDq4}, one can see that when we add the deformed boundary term
in this dilaton black hole background, a holographic flat band emerges as that in RN black hole.
In addition, the band is mildly dispersive for the small momentum. In the large momentum region,
the band is dispersionless\footnote{In Fig\ref{FlatBand3DDq4},
the band seems to disappear in large momentum region. We must point out that
it is the numerical artifact, which arises because in the large momentum region,
the peak becomes sharper and thinner. We can touch it by plotting the spectral function $A(\omega)$
as the function $\omega$ at fixed $k$.}.
By plotting the spectral function $A(\omega)$ as the function $\omega$ at fixed $k$
(large momentum region), we can find that the flat band located at $\omega\approx 3.237$,
which is approximately equal to the effective chemical potential $|\mu_{q}|$.
It is reasonable because the frequency is measured with respect to the effective chemical potential.

At the same time, a sharp quasiparticle-like peak also occurs near $k_{F}\approx 0.350920$.
We are also interested in the behavior of the spectral function
in the region of small $\tilde{k}=k-k_{F}$ and $\omega$ for the non-relativistic fermionic fixed point case.
By numerical analysis, one can fit the dispersion relation
as follow(Fig.\ref{DRHSFlatBandq4}):
\begin{eqnarray} \label{ScalingFBHeight}
\tilde{\omega}(\tilde{k})&\sim& \tilde{k}^{\delta},~~~\delta\approx 1.
\end{eqnarray}

We note that the dispersion relation is also linear in the case of
the non-relativistic fermionic fixed point as that of the relativistic fermionic fixed point case.
\begin{figure}
\center{
\includegraphics[scale=0.6]{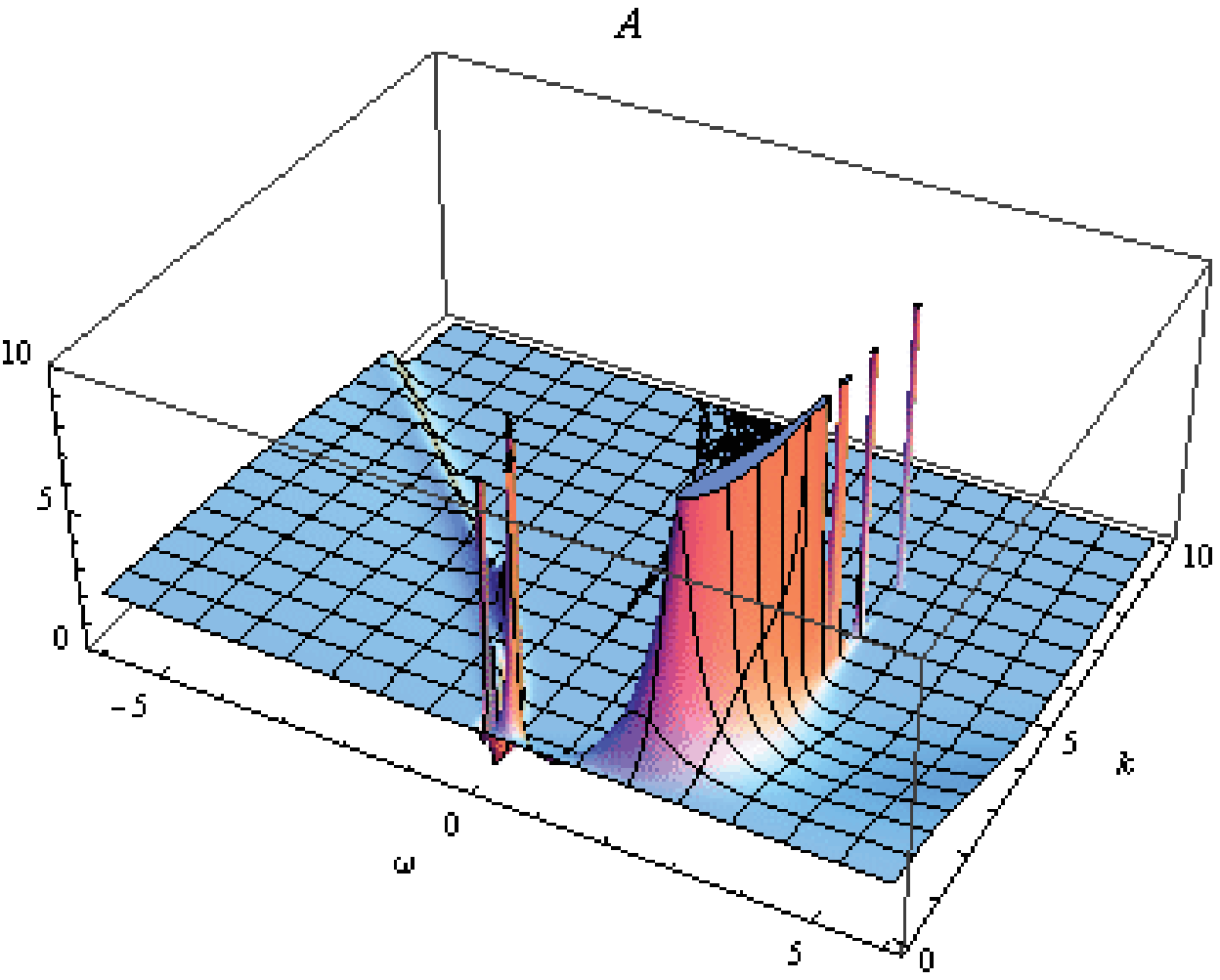}\hspace{1cm}
\includegraphics[scale=0.6]{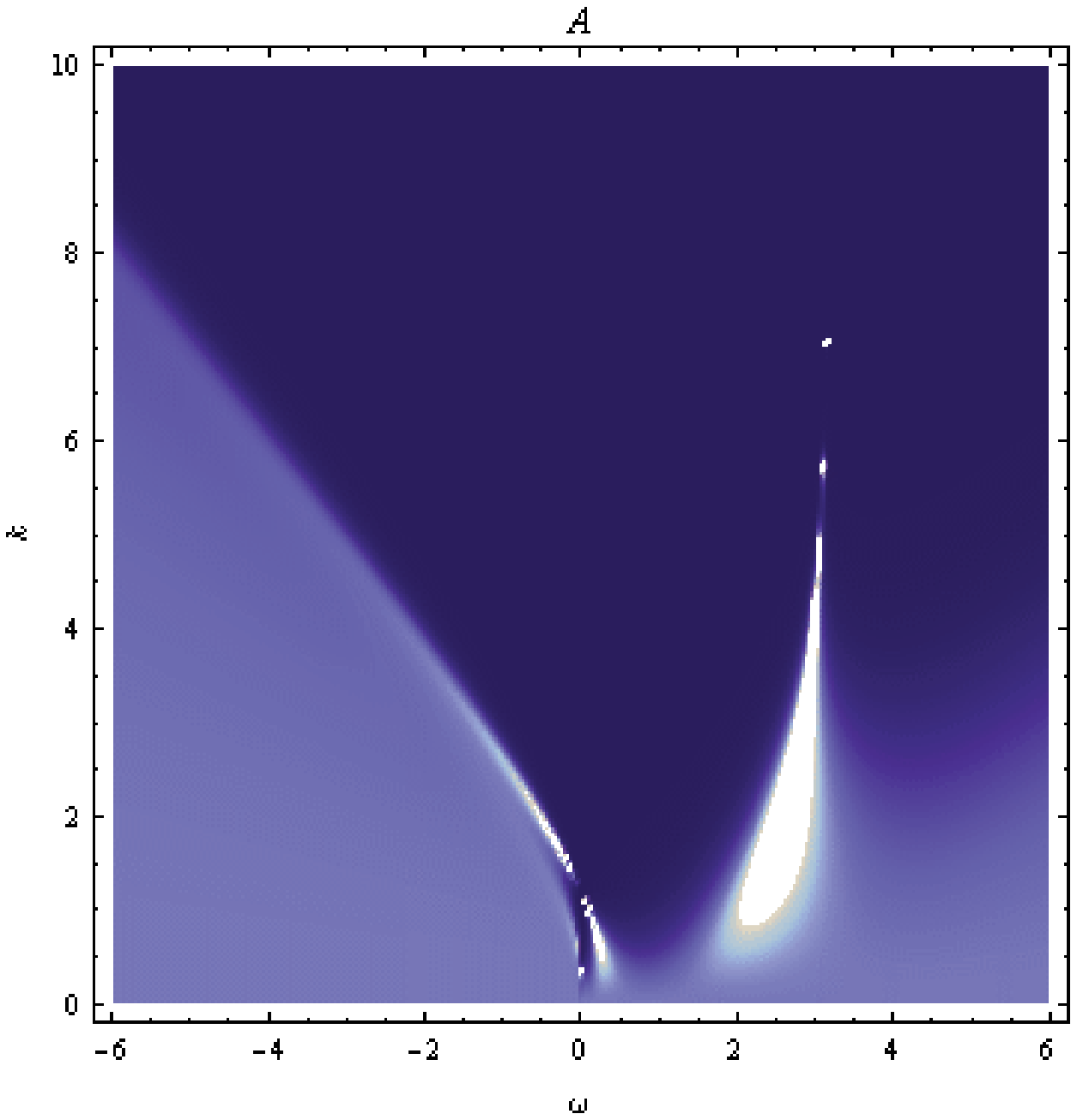}\\ \hspace{1cm}
\caption{\label{FlatBand3DDq4} The 3d and density plots of the spectral function $A(\omega,k)$
for $m=0$ and $q=4$ ($\mu_{q}\approx -3.37$). A holographic flat band emerges near $\omega\approx 3.237$,
which approximately equals to $|\mu_{q}|$. At the same time, a sharp quasiparticle-like peak also occurs near $k_{F}\approx 0.350920$.}}
\end{figure}
\begin{figure}
\center{
\includegraphics[scale=0.9]{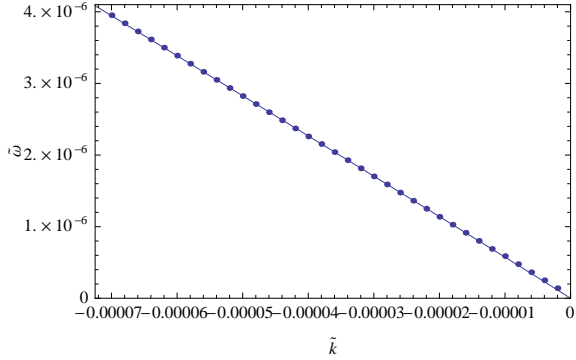}\\ \hspace{1cm}
\caption{\label{DRHSFlatBandq4}The dispersion relation between $\tilde{k}$ and $\tilde{\omega}$ for the case of non-relativistic fermionic fixed point.}}
\end{figure}

As point out in the case of relativistic fermion fixed point \cite{dipoleJPWu}, with the decrease of $q$,
the peak becomes smoother and wider and for small $q$, the Fermi sea disappears.
Therefor, we would also like to turn down the charge to $q=2$ to see what happen
when the Lorentz violating boundary term is added instead of the relativistic boundary term.
One find that the flat band emerges again in Fig.\ref{FlatBand3DDq4}.
However, the Fermi sea also disappears as the case of relativistic fermion fixed point.
Even for $q=0$, corresponding to the boundary effective chemical potential $\mu_{q}=0$,
which is decouple between the spinor field and gauge field, we also find that
a flat band emerges at $\omega=0$ (Fig.\ref{FlatBand3DDq0}).
It is comparable with the case of Schwarzchild \cite{HFlatBand}.
In summary, as in the background of RN-AdS black hole,
when a Lorentz violating boundary term is added to the Dirac action,
there still exists a holographic flat band in the background of dilaton black branes.
It seems that the emergence of flat band is robust in the case of non-relativistic fermionic fixed point.

\begin{figure}
\center{
\includegraphics[scale=0.6]{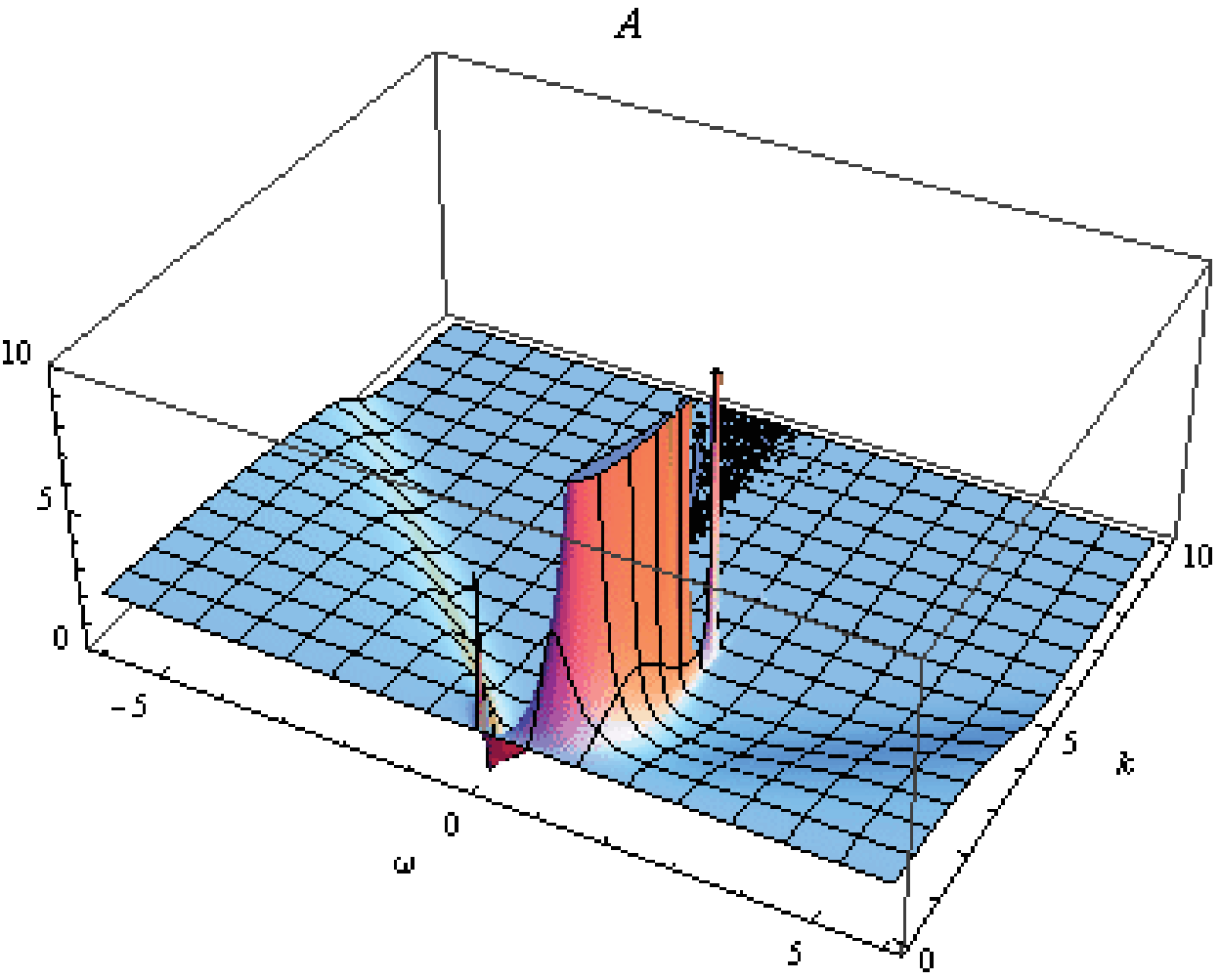}\hspace{1cm}
\includegraphics[scale=0.6]{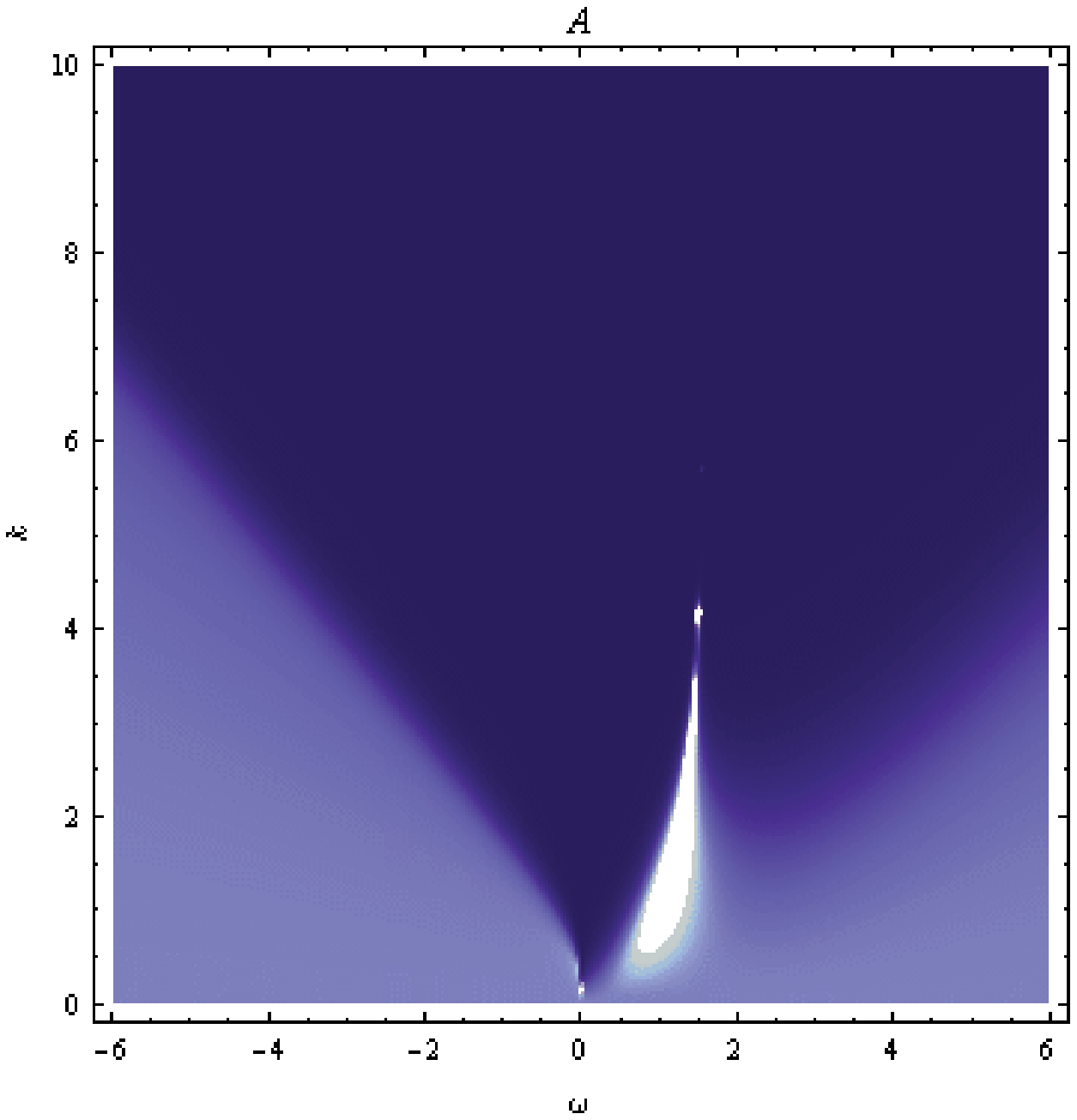}\\ \hspace{1cm}
\caption{\label{FlatBand3DDq4} The 3d and density plots of the spectral function $A(\omega,k)$
for $m=0$ and $q=2$ ($\mu_{q}\approx -1.686$). A holographic flat band emerges near $\omega\approx 1.62$,
which approximately equals to $|\mu_{q}|$.}}
\end{figure}
\begin{figure}
\center{
\includegraphics[scale=0.6]{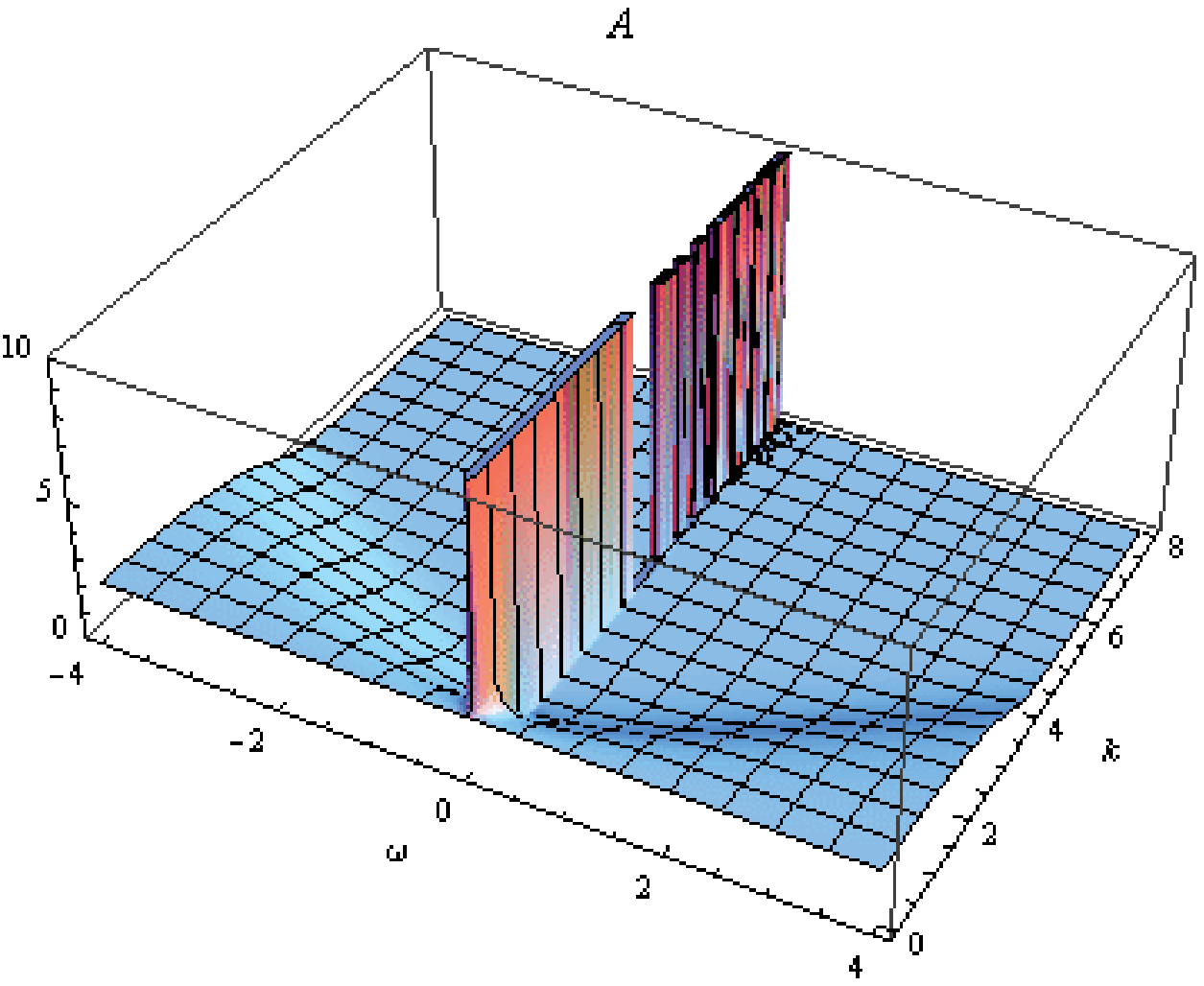}\hspace{1cm}
\includegraphics[scale=0.6]{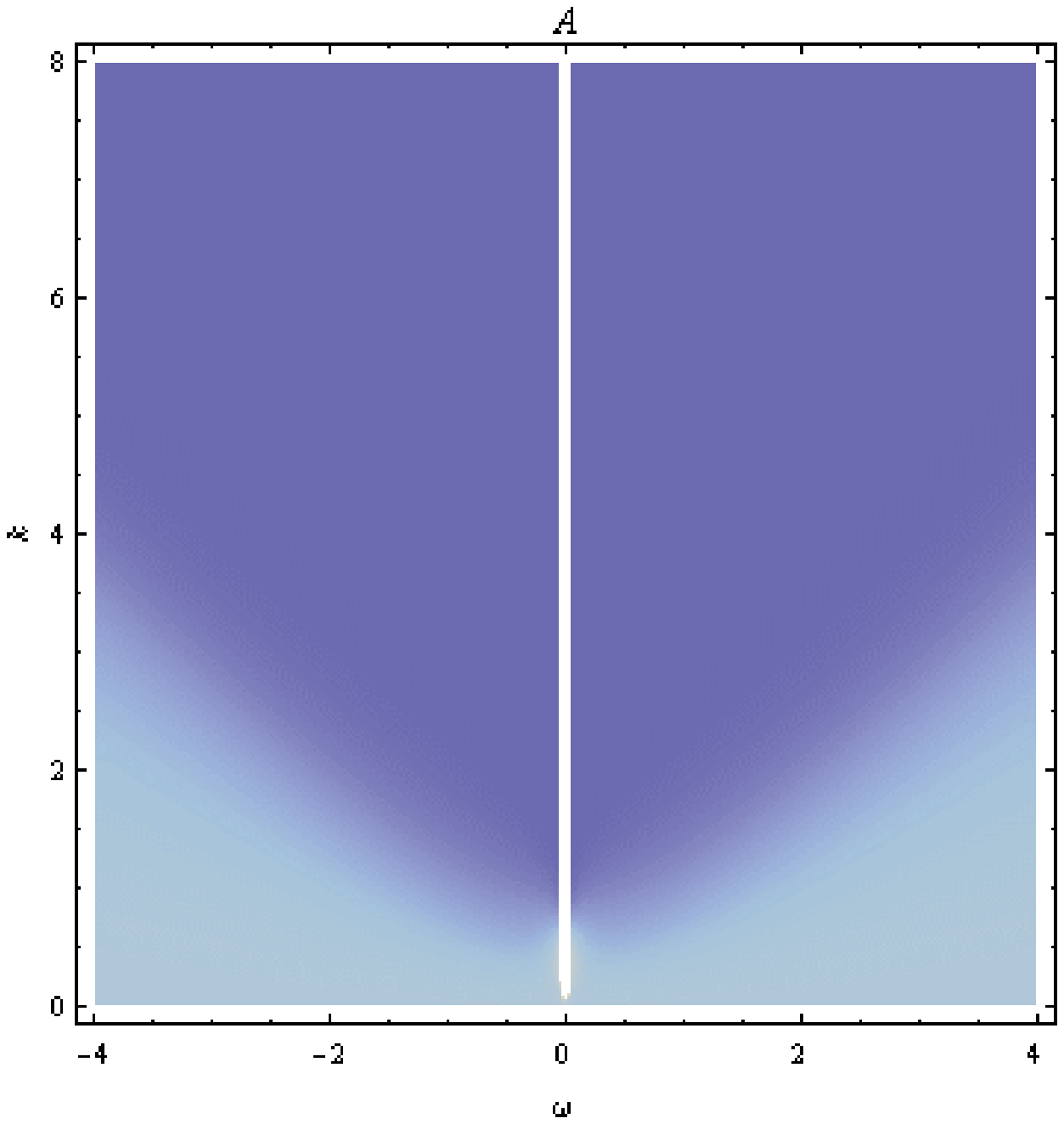}\\ \hspace{1cm}
\caption{\label{FlatBand3DDq0} The 3d and density plots of the spectral function $A(\omega,k)$
for $m=0$ and $q=0$ ($\mu_{q}=0$). A holographic flat band emerges at $\omega=0$.}}
\end{figure}

\section{Conclusion}

In this paper, we have investigated both the holographic relativistic and non-relativistic fermionic fixed points by imposing the relativistic boundary term and Lorentz violating that
in the dilatonic black brane with a Lifshitz like IR geometry and $AdS_4$ boundary.
On these two fixed points, by choosing proper parameters of the bulk fermion, we find
that there are emergent Fermi-surface structures. In addition, we also note that due to the presence of the holographic flat band, the Fermi momentum is suppressed in the case of non-relativistic fermionic fixed points, just like what happened in RN-AdS background\cite{HFlatBand}.
However, the low energy behavior both exhibits a linear dispersion relation on these two fixed points.
As pointed out that in Section \ref{SubS1}, the analytical understanding of the low energy behaviors of the dual relativistic fermionic system in the Einstein-Maxwell-dilaton black holes has been worked out in Ref.\cite{dilatonextention4}.
It will be interesting to give an analytical understanding of the low energy behaviors of the dual non-relativistic fermionic system in this charged dilaton black branes following the method developed in Ref.\cite{dilatonextention4}.

By the way, we also find there always exist a flat band without Landau level if one put on the Lorentz violating boundary term in this background. As the case in the RN-AdS background, the band is mildly dispersive for the low momenta. In the high momentum region, the band becomes sharper and dispersionless.
It seems that the emergence of flat band is robust in the case of non-relativistic fermionic fixed point.

\begin{acknowledgments}

We especially thank Hongbao Zhang for his collaboration in the related project.
J.P. Wu is partly supported by NSFC(No.10975017) and the Fundamental Research Funds for the central Universities.
W.J. Li is supported in part by the NSFC under grant
No.10605006 together with 10975016 and by the Fundamental Research Funds for the
Central Universities.

\end{acknowledgments}

\end{document}